\documentclass{article}

\usepackage{arxiv}
\usepackage{makecell}
\usepackage[utf8]{inputenc} 
\usepackage[T1]{fontenc}    
\usepackage{hyperref}       
\usepackage{color}
\usepackage[figuresright]{rotating}
\usepackage{url}            
\usepackage{booktabs}       
\usepackage{amsfonts}       
\usepackage{nicefrac}       
\usepackage{microtype}      
\usepackage{lipsum,amsmath}
\usepackage{graphicx}
\usepackage{natbib}
\usepackage{algpseudocode,algorithm,algorithmicx}
\setcitestyle{authoryear,open={(},close={)}}

\usepackage{setspace}
\title{Bootstrap Matching: a robust and efficient correction for non-random A/B test, and its applications}

\author{
  Zihao Zheng\\
  Department of Statistics\\
  University of Wisconsin-Madison\\
  Madison, WI, USA\\
  \texttt{zihao.zheng@wisc.edu} \\
  \And
  Carol Liu \thanks{To whom the corresponding should be addressed} \\
  Department of Marketing\\
  University of Wisconsin-Madison\\
  Madison, WI, USA\\
  \texttt{liu747@wisc.edu} \\
}

\begin{document}
\maketitle

\begin{abstract}
A/B testing, a widely used form of Randomized Controlled Trial (RCT), is a fundamental tool in business data analysis and experimental design. However, despite its intent to maintain randomness, A/B testing often faces challenges that compromise this randomness, leading to significant limitations in practice. In this study, we introduce Bootstrap Matching, an innovative approach that integrates Bootstrap resampling, Matching techniques, and high-dimensional hypothesis testing to address the shortcomings of A/B tests when true randomization is not achieved. Unlike traditional methods such as Difference-in-Differences (DID) and Propensity Score Matching (PSM), Bootstrap Matching is tailored for large-scale datasets, offering enhanced robustness and computational efficiency. We illustrate the effectiveness of this methodology through a real-world application in online advertising and further discuss its potential applications in digital marketing, empirical economics, clinical trials, and high-dimensional bioinformatics.
\end{abstract}

\keywords{\\Randomized Control Trial \and Bootstrap \and Matching \and Hypothesis Testing \and Online Advertising}

\section{Introduction}

A/B testing, a specific application of the Randomized Controlled Trial (RCT) methodology, has become a cornerstone in business analysis, particularly in the evaluation of treatment effects on key success metrics, such as conversion rates, user engagement, and revenue. For example, to assess the impact of a new feature on the number of online orders, a randomized controlled trial might be designed where subjects (e.g., customers or users) are randomly assigned to either the treatment or control group based on their IDs. In such scenarios, standard statistical techniques like t-test, Analysis of Variance (ANOVA), Analysis of Covariance (ANCOVA) etc are commonly employed to estimate the treatment effect, leveraging the robustness of t-tests and regression models against violations of normality assumptions. Even when the data deviates from a normal distribution, these methods remain valid, provided appropriate data pre-processing steps, such as log transformation and outlier detection, are applied.

However, the foundational assumption of randomness between the experimental and control groups is crucial for the validity of these statistical approaches. Randomization serves to eliminate biases and ensure that confounding variables are evenly distributed across treatment conditions, thus isolating the effect of the intervention. When this randomness is compromised, the integrity of the experimental design is undermined, and the risk of confounding increases, leading to potentially misleading conclusions.

Despite the theoretical simplicity of ensuring random assignment, practical challenges often arise that violate this assumption. For instance, a common yet flawed approach in A/B testing might involve assigning subjects to treatment and control groups based on the last digit of their IDs—e.g., all IDs ending in an odd number are assigned to the experimental group, while those ending in an even number are assigned to the control group. While seemingly arbitrary, this method is problematic because it assumes that the digit used for assignment is unrelated to any other factors influencing the outcome. In reality, if the same digit-based assignment rule is applied across multiple experiments or if the digit correlates with other variables, the assumption of randomness is violated, introducing bias and confounding into the analysis.

The importance of maintaining true randomization in RCTs and A/B testing cannot be overstated. Randomization is not just a procedural step; it is the bedrock upon which the validity of the experiment rests. Without it, the very essence of controlled experimentation is lost, and the results become suspect. As noted by Fisher \citep{fisher1966design}, the power of the RCT lies in its ability to "eliminate the effects of lurking variables" through random assignment. Modern research continues to emphasize the critical nature of this principle \citep{imbens2015causal, athey2017econometrics}, highlighting the need for rigorous experimental design to avoid the pitfalls of non-random assignment.

The challenge becomes even more pronounced when conducting observational studies, where running a controlled experiment is either impractical or impossible. In such cases, the goal is often to estimate the Average Treatment Effect on the Treated (ATT). However, unlike randomized controlled trials (RCTs), observational studies do not benefit from the random assignment of subjects to treatment and control groups. Consequently, the assumption of randomness between these groups is rarely, if ever, satisfied. This lack of randomization introduces significant complexities, as the treatment and control groups may differ systematically in ways that confound the estimation of the treatment effect. Such differences can arise from selection bias, where the factors that influence the assignment to treatment are correlated with the outcomes of interest, thereby violating the core assumptions of causal inference.

Addressing non-randomness in the pre-treatment period is crucial for ensuring that subsequent statistical inferences are robust and credible. One of the most widely adopted approaches to tackle this issue is matching, with Propensity Score Matching (PSM, \citep{rosenbaum1983central}) being among the most popular techniques. PSM involves estimating the probability that a subject is assigned to the treatment group based on observed covariates, known as the propensity score . Subjects in the treatment group are then matched with control group subjects who have similar propensity scores, thereby creating a pseudo-randomized sample. This process helps to mitigate confounding bias by ensuring that the treatment and control groups are comparable on the observed covariates, making the assumption of randomness more plausible in observational studies .

PSM is just one of many matching methods available. Other variants, such as nearest-neighbor matching \citep{rosenbaum1985constructing}, caliper matching \citep{cochran1973controlling}, and stratification on the propensity score, offer different trade-offs in terms of bias reduction and efficiency. Beyond propensity scores, more recent developments in matching techniques leverage machine learning algorithms to create more flexible and accurate matching procedures. For example, methods like genetic matching \citep{diamond2013genetic} use optimization algorithms to find the best balance between treated and control groups across multiple covariates, coarsened exact matching (CEM, \citep{iacus2012causal}) groups subjects into strata with identical or near-identical covariate values before matching, and CausalImpact \citep{brodersen2015inferring} takes the advantage of Bayesian structural time series predictions.

The use of matching, particularly in its various forms, is a powerful tool for mitigating the limitations of non-randomized studies. Yet, it is essential to carefully consider the choice of matching method based on the study context and the characteristics of the data to ensure that the resulting inference is both valid and reliable.

Though the idea is neat and often useful, there are at least three limitations for this idea:

\begin{itemize}
    \item \emph{Artificial Feature Space:} The feature space in which subjects are matched is often artificially defined. If the feature vector is not appropriately selected, the matching process may lead to overfitting, where the model captures noise rather than true underlying patterns. This overfitting can result in biased estimates of the treatment effect.
    \item \emph{Lack of Robustness in Treatment Group:} Traditional PSM and other matching methods do not comprehensively consider the robustness of the experimental group. For instance, in PSM, logistic regression is performed using all subjects, and every treated subject is matched with the most similar control subject. However, this does not account for the variability and potential outliers within the treatment group, which could lead to unreliable estimates.
    \item \emph{Computational Complexity:} The computational complexity of matching can be prohibitive, especially in large-scale studies. For example, when the number of treated subjects approaches $10^6$or more, the matching process becomes computationally expensive, even if parallel computing is employed. This limitation makes it difficult to apply matching methods to big data scenarios efficiently.

\end{itemize}

Motivated by these challenges and the limitations of current matching algorithms, we propose a "Bootstrap Matching" algorithm that balances the strengths of matching with the robustness of the bootstrap method. This approach aims to enhance the reliability of treatment effect estimates while addressing the computational challenges associated with large datasets. The details of the algorithm are discussed in the second section, followed by a real-world application in the third section.

\section{Methodology and Algorithm}

Let's denote our total subject size $n+m$ where $n$ subjects are in experimental group and $m$ subjects are in control group. $D_i, i\in\{1,2,\cdots, m+n\}$ be the group indicator and the response that we would like to evaluate to be $y$. Specifically, for each subject $i$, we observe $y_{i1, \cdots, y_{it}, y_{i(t+1)}, \cdots, y_{iT}}$ where the first $t$ numbers are observed before the treatment and the last $(T-t)$ numbers are observed after the treatment. Also, we have our feature matrix $X = [X_1', X_2', \cdots, X_{m+n}']'$ where each vector $X_{i}$ be the vector of length $k$ ($k$ features that we would like to evaluate).

The algorithm is described as the following:

\begin{algorithm}
\label{Bootstrap Mathcing algorithm}
  \caption{Bootstrap Matching algorithm}
  \hspace*{0.02in} 
  {\bf Input:}  Feature matrix $X$, group indicator $D_i, i\in \{1,2,\cdots, m+n\}$ and the response in the pre-period $y_{il}, i\in \{1,2,\cdots, m+n\}, l\in \{1,2,\cdots, t\}$, $N$ the number of bootstrap sampling replicates, $q$ the proportion of bootstrap sampling ratio.\\
  {\bf Output:} Effect size of treatment $\alpha$ and its false discovery rate $p$
  \begin{algorithmic}[1]
  \State Repeat from $n=1,2,\cdots, N$
  \State Bootstrap sample $(m+n)q$ subjects among all $(m+n)$ subjects and get all the corresponding statistics: $X_{B}, D_{B}, y_B$.
  \State Do matching (e.g., PSM) using only the bootstrap sample $X_B, D_B, y_B$ and get the matched data. Do statistical inference and report the estimated effect size $\alpha_n$ and p-value $p_n$.
  \State Calculate the average of estimated effect size, and the adjusted p-value or False Discovery Rate (FDR) $p$.
  \end{algorithmic}
\end{algorithm}

The parameters of the bootstrap method, particularly $q$ and $N$, can be adjusted based on factors such as sample size, computational efficiency, and other relevant considerations. As demonstrated in the examples below, the method remains robust when the combined sample size $m+n$ is large, which aligns with the primary use cases. A crucial and non-trivial aspect of this approach is how to combine the bootstrap estimates $\alpha_n$ and $p_n$ to obtain the final estimates $\alpha$ and $p$. Since $\alpha_n$ is known to be unbiased due to the construction of the bootstrap, taking the mean of $\alpha_n$ provides an accurate estimate of $\alpha$. However, greater care is needed when determining the significance level. Simply averaging the p-values $p_n$ is not advisable, as significance levels (and associated noise) depend on sample size, and the proposed approach involves repeated testing across $N$ iterations. Instead, a suitable multiple testing correction should be applied to appropriately adjust the p-values $p_n$, ensuring a more reliable overall significance estimate.

When dealing with multiple testing, it is crucial to adjust for the increased risk of Type I errors. Several established methods can be employed to address this topic. The FWER is the probability of making at least one Type I error among all the hypotheses tested \citep{shaffer1995multiple}. The Bonferroni correction is a commonly used method for controlling FWER, which simply divides the desired significance level by the number of tests. While straightforward, it can be overly conservative, particularly when dealing with a large number of tests, leading to a reduction in statistical power. Benjamini–Hochberg (BH) method \citep{benjamini1995controlling} controls the False Discovery Rate (FDR), which is the expected proportion of false positives among the rejected hypotheses. The BH method ranks the p-values and compares them to an increasing threshold, allowing more discoveries while controlling for false positives. It is widely used in high-dimensional data settings, such as genomics, where large numbers of hypotheses are tested simultaneously. An extension of the BH method, Storey’s q-value approach \citep{storey2002direct} also controls FDR but estimates the proportion of true null hypotheses to achieve more accurate control. This method is particularly useful when the data contains many true null hypotheses, as it provides a less conservative adjustment compared to the traditional BH method. 

All the approaches mentioned above provide a corrected version of p-value which controls the global type oen error. As p-value by definition is the probability of data given null hypothesis, doing average does not provide good interpretations. On the other hand, the Local False Discovery Rate (LFDR) is a powerful and flexible approach to multiple hypothesis testing that extends the concept of the False Discovery Rate (FDR) and provides interpretability when doing average. While the FDR controls the expected proportion of false positives among all rejected hypotheses, the LFDR specifically focuses on the probability that a particular hypothesis is a false positive, given its test statistic. The LFDR was first introduced by Dr. Efron, who provided a statistical framework for estimating LFDR using empirical Bayes methods. \cite{efron2001empirical} proposed estimating the LFDR by modeling the distribution of test statistics and leveraging the mixture of null and alternative distributions to derive posterior probabilities. This method is particularly useful when the exact distribution of test statistics under the null hypothesis is complex or unknown. Efron's work laid the groundwork for further advancements in LFDR methodology. For instance, Matthew Stephens introduced the Adaptive Shrinkage (ASH) method \citep{stephens2017false}, which provides a flexible and efficient way to estimate LFDR by shrinking test statistics toward zero using empirical Bayes techniques and \cite{zheng2021mixtwice} developed the MixTwice method, a novel approach to estimating LFDR that leverages a two-stage mixture model. The final estimated $p$ can be calculated by average LFDR, corrected from $p_i$.

\section{Examples and Applications}

\subsection{Real data example in online advertising}

In this section, we present an evaluation of the bootstrap matching algorithm applied to a real-world example where the target metric is the number of orders in an advertising setting. The experiment in question was designed using a digit-tail rule, where all IDs ending in 0 were assigned to the treated group, with the remaining IDs forming the control group. This setup exemplifies a typical non-random design in A/B testing, where the randomness assumption is violated, potentially leading to biased results.

To thoroughly assess the performance of the bootstrap matching approach, we selected a 12-day period around the treatment, consisting of 6 days before and 6 days after the treatment was administered. The total sample size for this study was approximately 400,000 observations, providing a substantial dataset for analysis.

\begin{figure} 
    \centering
    \includegraphics[width=0.5\columnwidth]{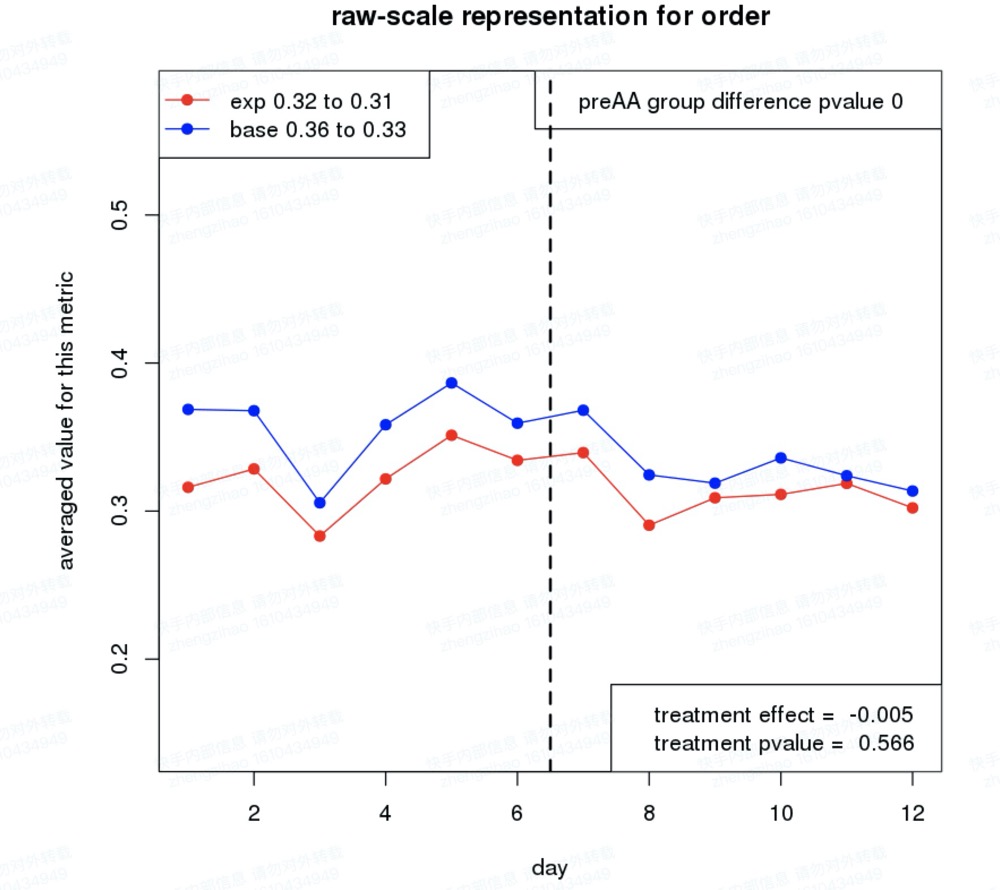}
    \caption{\textbf{How the pre-period randomness is violated?} This figure shows the average of resposne (number of orders) in each day. Difference colors indicate experimental group and control}
    \label{fig1}
\end{figure}

The impact of the non-random design is evident in Figure~\ref{fig1}, which illustrates the violation of the randomness assumption during the pre-treatment period. Specifically, the p-value for the pre-period comparison is less than 0.01, indicating a significant imbalance between the treatment and control groups before the intervention. This imbalance poses a serious threat to the validity of traditional A/B test analyses, as it suggests that differences observed in the post-treatment period may be driven by pre-existing differences rather than the treatment itself.

To address this issue, we implemented the bootstrap matching algorithm with the bootstrap parameter set to $N = 300$. In each bootstrap iteration, we randomly sampled 10,000 observations from the dataset to perform the matching. The results, as shown in Figure~\ref{fig2}, demonstrate the effectiveness of the proposed method in achieving balance between the treatment and control groups during the pre-treatment period. The figure highlights how the bootstrap matching algorithm reduces the pre-period imbalance, thereby enhancing the validity of subsequent causal inferences.

However, it is important to note that, despite the overall success of the method, some bootstrap samples still exhibit substantial imbalance between the groups. This occurrence underscores the inherent non-robustness of traditional matching techniques, which may fail to achieve balance in certain scenarios. The strength of the bootstrap matching algorithm lies in its ability to mitigate this issue by performing a large number of replicates, thus averaging out the effects of any individual imbalanced sample. This replicative approach not only improves the reliability of the matching process but also provides a more robust framework for causal inference in non-randomized settings.

\begin{figure} 
    \centering
    \includegraphics[width=0.5\columnwidth]{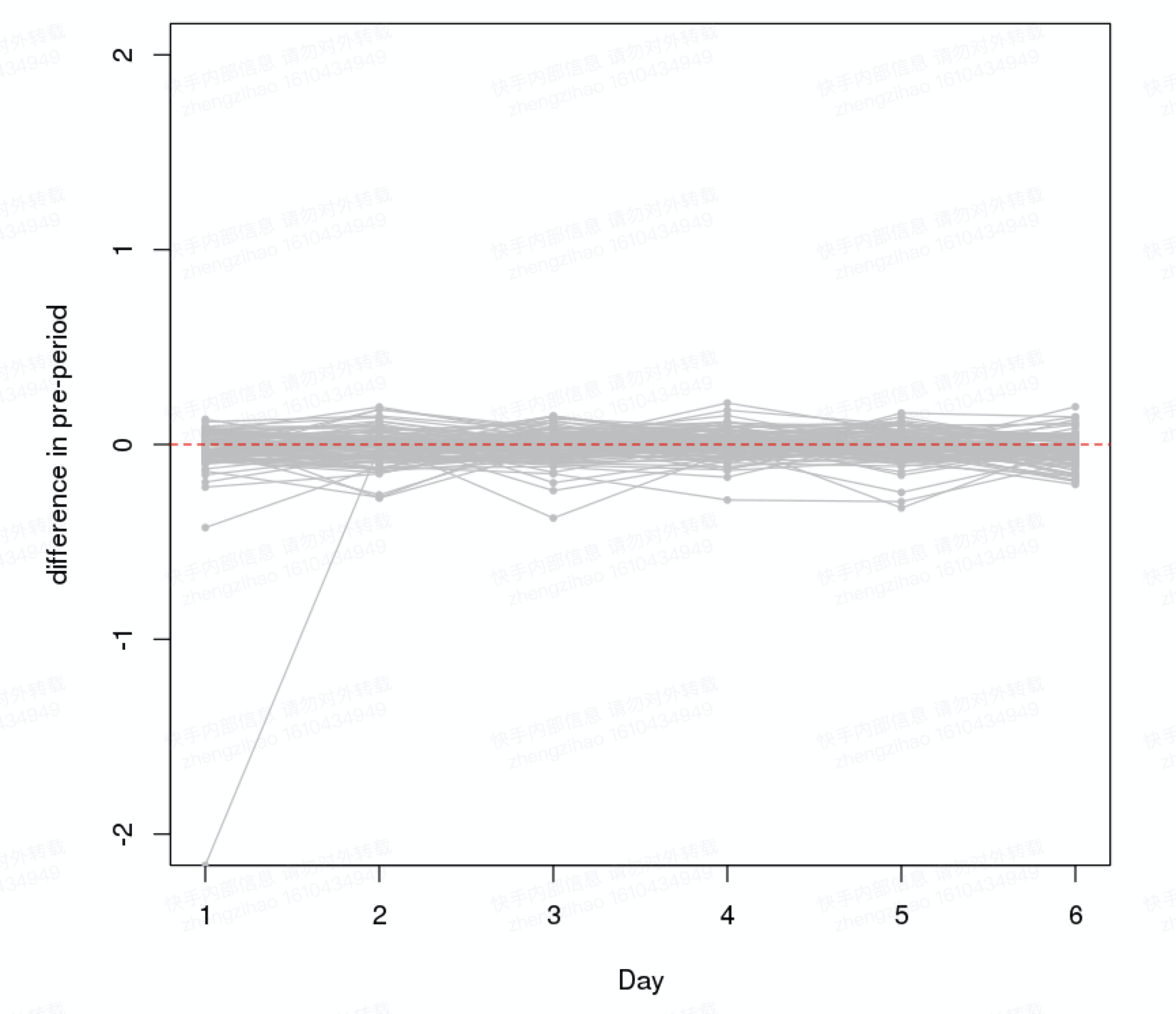}
    \caption{\textbf{The benefit of bootstrap matching.} After bootstrap matching, the difference in the pre-perior was plotted over time, for each bootstrapping sample.}
    \label{fig2}
\end{figure}

\subsection{Generalized applications}

Beyond the example of online advertising using a digit-tail rule with less rigorous diversion configurations, the proposed bootstrap matching approach has broader applications across various domains. These include digital marketing, high-dimensional bioinformatics, empirical economics, and clinical trials, among others. The versatility of this method makes it well-suited for addressing non-randomness and confounding factors in a wide range of scenarios where traditional randomized controlled trials (RCTs) may not be feasible or effective.

For example, one particularly relevant application is in observational studies conducted within Google Search Labs. Unlike controlled experiments where participants are randomly assigned to treatment or control groups, observational studies in this context often involve users who self-select into new features or experiences based on their personal preferences, behaviors, or demographics. For example, users who choose to test a new search feature might be more tech-savvy, frequent users, or have different search patterns compared to the general population. This self-selection introduces a bias, as the individuals who opt-in to new features may differ systematically from those who do not.

In the field of online advertising, Advertiser Experiment (AE) with budget control settings presents unique challenges, particularly when it comes to maintaining randomization. While many experiments, such as those conducted by Google, typically randomize at the user or cookie level—a strategy that is generally effective for large sample sizes—AE often requires more nuanced approaches to ensure randomness. A key issue arises when multiple advertisers target the same population, such as users or keywords, within a shared platform. In these cases, advertisers might allocate a common budget across different campaigns. This shared budget creates a potential spillover effect between the treatment and control groups, where the behavior or performance of one group could influence the other. While various experimental designs, such as cluster randomization \citep{donner2000design}, have been proposed to address these challenges, these methods often involve complex setups. In contrast, bootstrap matching offers a simpler and more flexible post-randomization solution. 

\begin{figure} 
    \centering
    \includegraphics[width=0.5\columnwidth]{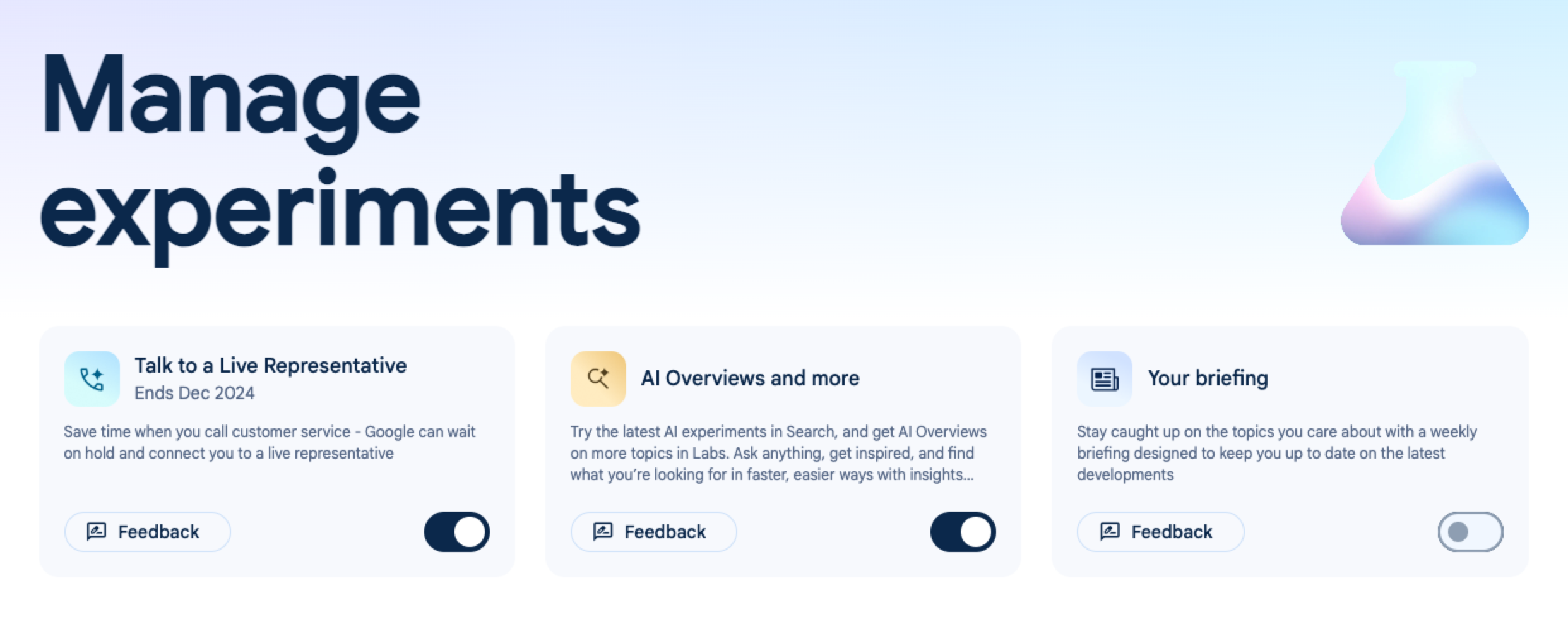}
    \caption{\textbf{Google Labs}}
    \label{fig2}
\end{figure}

Bootstrap matching is not only advantageous in the context of online advertising and digital marketing but also holds significant potential in high-dimensional bioinformatics, social science, and clinical trials. In high-dimensional bioinformatics, researchers often deal with datasets that contain thousands or even millions of variables, such as gene expression levels \citep{audic1997significance,raser2005noise,anders2010differential,clough2016gene}, genomic sequences \citep{church1984genomic,pareek2011sequencing}, and high dimensional peptide arrays \citep{zheng2020disordered,mergaert2021divergent,amjadi2024novel,parker2022novel,parker2024novel}. Traditional matching methods can be enhanced by bootstrap matching, which allows for more robust inferences by repeatedly resampling and matching subsets of data. This approach helps in controlling for confounding factors and improving the accuracy.

In the area social science research and empirical economics, observational studies frequently grapple with the issue of non-random treatment assignment, where subjects self-select into treatment or control groups based on unobserved characteristics. This non-randomness can introduce biases that compromise the validity of causal inferences \citep{altonji2005selection, heckman2015causal,athey2017state, angrist2010credibility,liu2022progressive}. Bootstrap matching provides a valuable tool for mitigating these biases by generating multiple matched samples and averaging results across them, thus reducing the influence of any single, non-representative match. This approach is particularly useful in studies where randomization is not feasible due to ethical or logistical constraints.

Clinical trials, especially in the field of personalized medicine, often face challenges related to the heterogeneity of patient populations. When randomized controlled trials (RCTs) are not possible or ethical, observational studies become necessary \citep{concato2000randomized,d2007propensity,hernan2016using}. Bootstrap matching offers a robust alternative by simulating the randomization process through repeated resampling and matching, which helps to balance treatment and control groups on a multitude of covariates. This is particularly beneficial in early-phase clinical trials, where sample sizes are typically small, and in post-market surveillance studies, where the goal is to assess the real-world effectiveness of interventions across diverse patient populations.

\section{Conclusions and Future directions}

In conclusion, Bootstrap Matching presents a versatile and robust framework for addressing the challenges associated with non-randomized experimental designs and observational studies. By combining the principles of bootstrap resampling with matching techniques, this method enhances the reliability of causal inferences, particularly in situations where traditional randomized controlled trials are impractical or impossible to implement. The flexibility of Bootstrap Matching allows it to be adapted to a wide range of applications, from online advertising and digital marketing to high-dimensional bioinformatics and clinical trials.

Moreover, by mitigating issues such as overfitting, lack of robustness, and computational complexity inherent in standard matching algorithms, Bootstrap Matching offers a practical solution for large-scale data analysis. The real-world examples discussed, including the digit-tail rule in online advertising and observational studies in various domains, underscore the effectiveness of this approach. As data sets continue to grow in size and complexity, Bootstrap Matching is poised to become an increasingly valuable tool for researchers and practitioners seeking to derive meaningful insights from non-randomized data.

Looking ahead, there are several promising avenues for further research and development in the context of Bootstrap Matching. One important direction involves enhancing the computational efficiency of the algorithm, particularly when dealing with ultra-large datasets that are becoming increasingly common in fields like genomics and digital marketing. While the current method leverages parallel computing to some extent, more sophisticated techniques, such as distributed computing frameworks or GPU acceleration, could be explored to further reduce computation time and increase scalability.

Another potential area of future exploration lies in refining the matching criteria used within the Bootstrap Matching framework. As the landscape of data science continues to evolve, integrating machine learning models to improve the accuracy and relevance of the matching process could yield more precise results. Additionally, extending Bootstrap Matching to handle time-varying treatments or longitudinal data could open new possibilities in the analysis of complex experimental designs, particularly in social sciences and clinical trials where such data structures are prevalent.

\clearpage
\bibliography{references}

\end{document}